\documentclass[11pt,twoside]{article}
\usepackage{asp2010}

\resetcounters

\begin{document}

\title{Clumpology of Starbursts in the WiggleZ Dark Energy Survey}
\author{Emily~Wisnioski, Karl~Glazebrook, Chris~Blake}
\affil{Centre for Astrophysics and Supercomputing, Swinburne University of Technology, P.O. Box 218, Hawthorn, VIC 3122, Australia (ewisnios@astro.swin.edu.au)}
\author{\& The WiggleZ Team}

\begin{abstract}
We have observed H$\alpha$ emission from a sample of 13 star-forming galaxies from the WiggleZ Dark Energy Survey in the redshift range $z\sim1.2-1.5$ and stellar mass range $9.8<\log$(M$_*$/M$_{\odot}$)$<11.6$ with the integral field spectrograph OSIRIS on Keck, taking advantage of Laser Guide Star Adaptive Optics.  We detect multiple emission, 1--2 kpc size sub-components, or `clumps' within the H$\alpha$ spatial emission in 4 galaxies, extended emission in 4 galaxies and compact spheroidal emission in 5 galaxies. Considering our results in the context of unstable disk formation and a merger sequence, we find evidence of ordered orbital motion in the majority of galaxies as would be found in unstable gaseous disks.
\end{abstract}

\section{Morphologies at High Redshift}
Morphological studies made possible by deep imaging surveys have provided the first clues to the processes of high-redshift galaxy formation. It has become clear from these results that high-redshift galaxies exhibit different morphologies to local galaxies, with higher fractions of both compact and irregular morphologies \citep{Glazebrook:1995vn, 1996MNRAS.279L..47A}.  Initial interpretations of the complex structures seen in the Hubble Ultra Deep Field favor higher rates of major mergers, a natural consequence of hierarchical merging models in a $\Lambda$CDM Universe (e.g., \citealt{2003AJ....126.1183C, 2006ApJ...636..592L}), however, additional observations of irregular galaxies reveal properties reminiscent of local disk galaxies \citep{2005ApJ...627..632E}. Yet, unlike local disks, the galaxies at high redshift contain large dense regions or `clumps' of star formation, comparable to HII regions but of much larger size \citep{Jones:2010uf}. Models have been developed to explain the clumpy structures in disks and their evolution into galaxies with morphologies seen at $z=0$ (e.g., \citealt{1999ApJ...514...77N, 2004A&A...413..547I, 2007ApJ...670..237B, Elmegreen:2008fk}). 

Morphological studies alone do not provide enough information to distinguish between these kinematically distinct theories of galaxy formation. Spatially resolved spectroscopy is therefore essential for disentangling kinematics and star formation histories of these complex structures at high redshift. 

\section{Kiloparsec scale Observations of Star-Forming Galaxies at $z>1$}

The past five years have seen a great number of integral field spectroscopy (IFS) studies of nebular emission line H$\alpha$ in $z>1.5$ star-forming galaxies (e.g., \citealt{2006Natur.442..786G, 2009ApJ...697.2057L,2009ApJ...699..421W, 2009A&A...504..789E, 2008A&A...479...67N}). These redshifts are favored for the large number of Lyman break galaxies (LBGs) and sub-millimeter galaxies (SMGs) being discovered by methods tailored for high redshift and for the accessibility of H$\alpha$ in the near infrared.

A main focus of IFS studies has been the classification of galaxies kinematically to define clear observational distinctions between disks and interacting systems. Mergers observed with IFS \citep{2008A&A...479...67N,2009ApJ...699..421W} have been found to show clear kinematic steps across one resolution element indicating a velocity separation between two kinematically separate systems. Some galaxies also exhibit rotational signatures across IFS results and are well fit by disk models \citep{2006Natur.442..786G, 2009A&A...504..789E, 2010MNRAS.402.2291L}. Yet rigidly classifying galaxies in these two distinct groups is an over-simplification. Many of the samples of $>5$ galaxies present a variety of kinematic types including dispersion dominated systems, stable disks, unstable disks, minor mergers, merger, and merger remnants with definitions of each class varying from one sample to another. 

\section{Discovering extreme objects at $z>1$ using the WiggleZ Dark Energy Survey}

\begin{figure}[!h]
\includegraphics[scale=0.68,viewport=-40 10 600 400,clip]{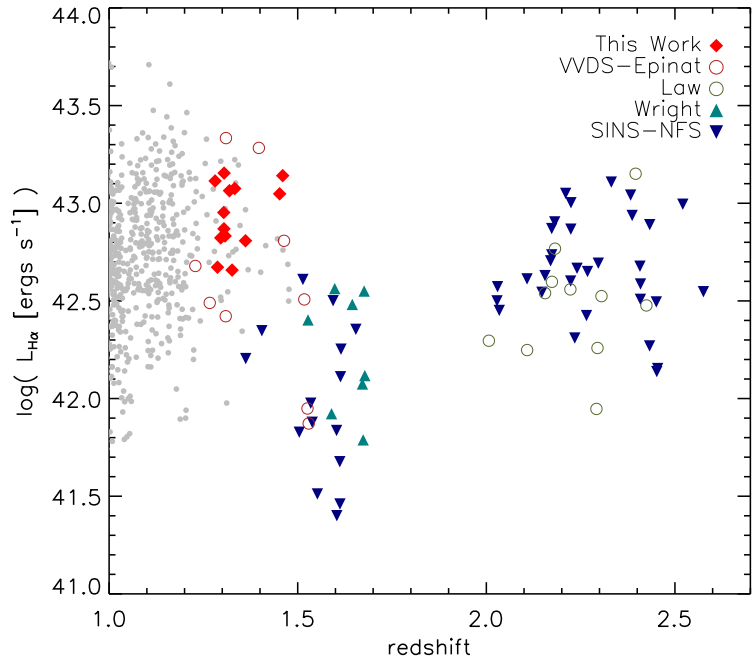}
\caption{
H$\alpha$ luminosity as a function of redshift for the largest high-redshift IFS surveys with our work shown by the red diamonds. The gray points show the [OII]~luminosity of other WiggleZ galaxies considered for this sample. The remaining symbols are from \cite{2009A&A...504..789E, 2009ApJ...697.2057L,2009ApJ...706.1364F,2009ApJ...699..421W}. 
\label{fig.litcomp}
}
\end{figure}

We present integral field spectroscopy of H$\alpha$ emission in 13 star-forming galaxies selected from the WiggleZ Dark Energy Survey at $z\sim1.3$ \citep{Drinkwater:2010bx}. By moving to slightly lower redshifts than previous studies of LBGs and SMGs we are able to study the internal kinematics of galaxies of comparable intrinsic luminosities (Figure~\ref{fig.litcomp}) but achieving higher spatial resolution and comparable signal to noise in a much shorter exposure time. WiggleZ galaxies are UV selected with color cuts to select the bluest galaxies across the whole redshift range of the survey. While the WiggleZ redshift distribution peaks at $z\sim0.6$, there exists a long thin tail of high-redshift galaxies that reaches as high as $z=1.5$.  These galaxies are identified from strong [OII]~emission and have star formation rates of 50--500~M$_{\odot}$\ yr$^{-1}$. There are about 100 such objects in the survey. The WiggleZ survey has made the current study of intensely star-forming galaxies at $1<z<2$ possible by finding rare starbursts that would otherwise be missed. These objects may be the link needed to bridge the gap between observations of local LIRGs/ULIRGs and high-redshift LBGs and SMGs at one of the most important epochs of mass assembly. 

\section{Results}
\begin{figure}[!h]
\includegraphics[scale=0.68,viewport=-30 200 600 760,clip]{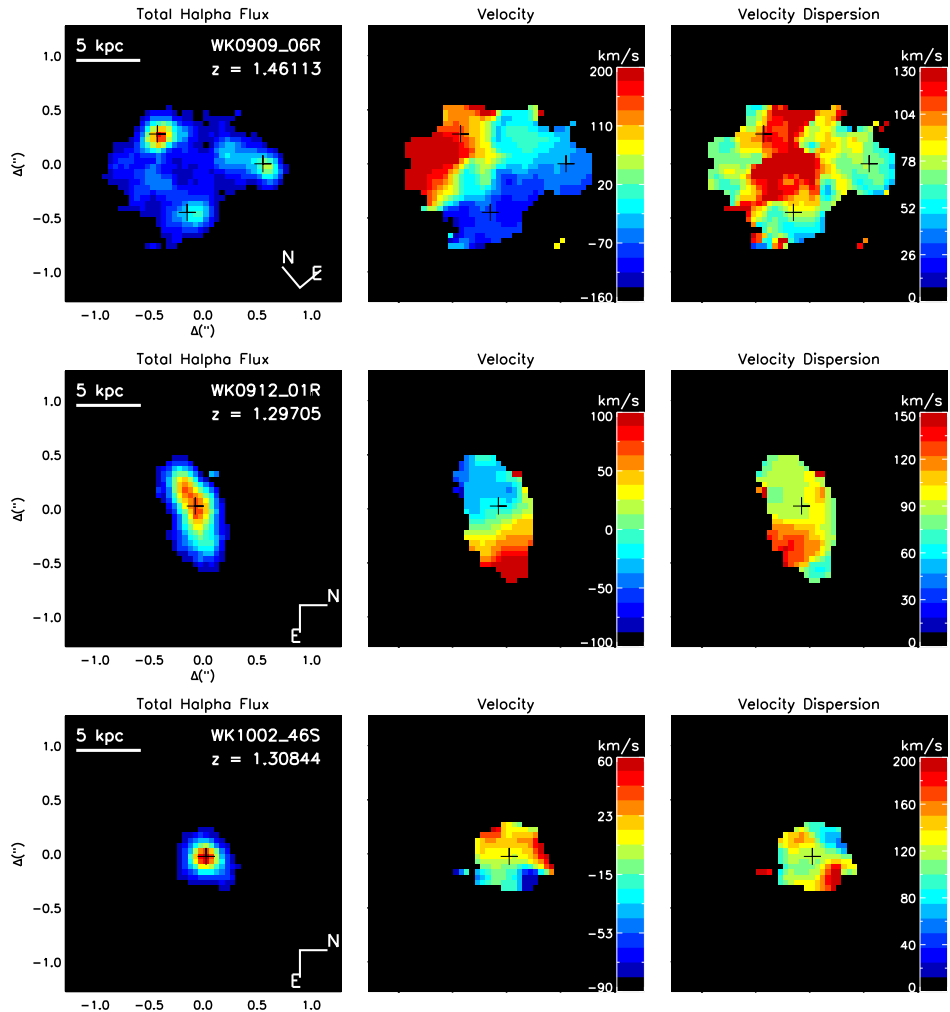}
\caption{
OSIRIS H$\alpha$ kinematics of 3 out of 13 galaxies in our sample, WK0909\_06R is an example of the multiple emission galaxies, WK0912\_01R is an example of the extended emission galaxies, and WK1002\_42S is an example of the single emission galaxies. {\it Left:} H$\alpha$ flux map, {\it middle} velocity map from the systemic velocity, {\it right} velocity dispersion map. The black crosses mark the peak(s) in H$\alpha$ flux. Velocity and velocity dispersion are measured in km/s. 
\label{fig.kinematics}
}
\end{figure}

We observed 13 star-forming galaxies using OSIRIS (OH Suppressing InfraRed Imaging Spectrograph) with the Keck II  laser guide star adaptive optics (LGSAO) system. By using an IFU we obtain an H$\alpha$ spectrum for every spatial pixel (spaxel) of the galaxy image.  A Gaussian profile is fitted to the H$\alpha$ emission in each spaxel of the cube to map out the H$\alpha$ flux distribution of the galaxy. A velocity map is constructed for each galaxy using the velocity shifts of the emission in each spaxel relative to the galaxy redshift. Velocity dispersion is measured in each spaxel from the r.m.s. of the fitted Gaussian line profile corrected for instrumental broadening (R=3600). The kinematic maps are shown for three galaxies in Figure~\ref{fig.kinematics}.

We classify the 13 galaxies into three classes based solely on their morphologies; single emission: showing a single resolved source of H$\alpha$ emission, extended emission: showing multiple unresolved or extended components, and multiple emission: having multiple resolved regions of H$\alpha$ emission. These classifications are used to discuss the properties of galaxies with like morphologies in the context of different formation scenarios.
 
{\it The Clumpy Disk Hypothesis:  }
We investigate the scenario that the classes represent different stages of the clumpy disk models of \cite{Elmegreen:2008fk} simulations (EBE08). In these simulations gravitational instabilities result in multiple star-forming complexes forming under Jeans collapse within a young gas rich disk. As turbulent speeds decrease relative to the rotational speed of the disk the clumps migrate towards the center on a timescale of $\sim1$ Gyr. Finally, the clumps coalesce and form a slowly rotating pseudo-bulge \citep{1999ApJ...514...77N,2007ApJ...670..237B,Elmegreen:2008fk,2009arXiv0901.2458D, Krumholz:2010fk}.

The morphology of multiple emission systems is well matched to the early stages of clumpy disk formation ($t=175-450$ Myr) of the EBE08 simulations. The velocity of the clumps follows the global velocity structure of the galaxy, consistent with models \citep{2004ApJ...611...20I}. This is seen in WK0909\_06R (Figure~\ref{fig.kinematics}), where the major axis of rotation cuts across multiple clumps revealing no corresponding deviation in the rotation profile. The morphologies of the extended emission galaxies are either consistent with the system being observed at the clump coalescence phase ($t=600-800$ Myr) or the result of projection effects, with some galaxies well fit by disk models. Compact single emission galaxies have morphologies well matched to the later stage of simulated formation when the clumps have collapsed to form a bulge ($t=1000$ Myr). The high dispersions and masses of these systems are consistent with EBE08 bulges, predicted to become slow rotators with high turbulence of $\sigma_{los}\sim130$ km s$^{-1}$ in the inner regions with $M_{*}>10^{10}$ M$_{\odot}$.

Star-forming regions within a disk should have similar or consistent metallicities as they are evolving within the same conditions \citep{2004A&A...413..547I}  In a major or minor merger no similar correlation is expected in the metallicities of each galaxy. By integrating the light of each clump the clump metallicities are calculated from the H$\alpha$/[NII]~ratio. Metallicities of clumps in multiple emission galaxies, such as WK0909\_06R (Figure~\ref{fig.kinematics}) are equal within the errors, as predicted for the early stages of disk fragmentation. 
 
{\it The Merger Hypothesis:  }
However, with separations of 3--10 kpc the H$\alpha$ complexes may be sites of multiple regions of star formation merging from different parent galaxies. Without deep imaging of diffuse continuum light surrounding the H$\alpha$ emission to reveal tidal features characteristic of mergers, we need to utilize kinematic quantities to find support for this hypothesis. We analyze here a scenario in which our sample form a merger sequence with pre, ongoing, and post mergers. In this scenario the multiple emission galaxies are the initial stages of mergers when star formation in separate galaxies can be spatially differentiated. The extended emission galaxies are the next stage in the sequence representing ongoing merging activity (or pre-mergers with component confusion due to projection effects). The final stage in the proposed merger sequence is the single emission galaxies which represent merger remnants with high dispersions and star formation rates.

Although some of the galaxies in our sample show indications of ordered rotation indicative of disks from their velocity maps, none of the systems show a clear peak in velocity dispersion corresponding to the turnover of the velocity distribution as expected for disk galaxies. Although this may be explained by sensitivity effects, the lack of a peak in velocity dispersion, a feature observed in IFS studies in both local \citep{2002MNRAS.329..513D} and high redshift \citep{2006Natur.442..786G, 2009ApJ...706.1364F} disk galaxies, may indicate that we are seeing separate kinematic components merging. In this case, the observed peaks in dispersion may correspond to the centers of individual galaxies merging or be a result of shock fronts between merging systems. The majority of IFS results that have been classified as mergers show velocity maps where the velocity does not vary smoothly across the source but rather show a quick turnover within a resolution element or less \citep{2008A&A...479...67N, 2009ApJ...699..421W, 2009ApJ...697.2057L}. This step function is not present in the kinematics in our sample.

{\it Findings: }
Whilst it is natural to attempt to identify kinematic observations as disks or mergers, this simplified picture of galaxies becomes increasingly fallible at high-redshift. We are hesitant to split our data as has been done in previous studies (disk, merger, dispersion dominated, etc.) on an individual basis as we find that galaxies do not populate homogeneous classes but rather form a heterogeneous sample.  In a sample of 13 galaxies it is likely a variety of mechanisms are at work, even if disk instability is a dominant mechanism in star formation histories at high redshift, as our data supports, the disks are likely to undergo mergers in their lifetime further complicating kinematic interpretations. Future deep optical and infrared imaging will help to clarify the extended morphologies of these systems and assist in discriminating between turbulent disks and ongoing major/minor mergers.

\end{document}